# A BACKSTEPPING CONTROL METHOD FOR A NONLINEAR PROCESS - TWO COUPLED-TANKS

Vasile CALOFIR[1], Valentin TANASA[2], Ioana FAGARASAN[3],
Iulia STAMATESCU[4], Nicoleta ARGHIRA[5], Grigore STAMATESCU[6]

**Abstract:** *The aim of this work is to compute a level backstepping control strategy for a coupled tanks system. The coupled tanks plant is a component included in the water treatment system of power plants. The nonlinear-model of the process was designed and implemented in Matlab-Simulink. The advantages of the control method proposed is that it takes into consideration the nonlinearity which can be useful for stabilization and a larger operating point with specified performances. The backstepping control method is computed using the nonlinear model of the system and the performance was validated on the physical plant.*

## 1. Introduction

The coupled tanks system is a component included in the water treatment plant that is an important part of any power plant especially with the development of smart grids and with the stress of the enviromental issues nowadays. In traditional power plants (thermal, hydro, nuclear etc.) the water treatment part has the rol to    reduce and even eliminate polluting particles such as S, NO, $NO_2$, etc.. The level control of the used tanks has to be very precise and efficient . Many processes met in the water treatment / purification station can be modeled as systems with three open reservoirs.

To design a controller for maintaining constant level in such tanks, the need of a mathematical model of the plant is required [3],[4]. To obtain the mathematical model of the controlled process may be accomplished by analytical and experimental techniques. In case of the use of analytical models, mathematical models are obtained by applying the laws that describe functioning of the process (energy conservation laws, mass conservation law, etc.) taking into account the particularities of each process. In general, the mathematical models obtained by means of analytical designs are complex and most often contains nonlinear dependencies of the variables.

[1][3][4][5][6] Department of Automatic Control and Industrial Informatics, Politehnica University of Bucharest  vasile@shiva.pub.ro[1]; [3]ioana@shiva.pub.ro; [4]iulia.dumitru@aii.pub.ro;
[5]nicoleta.arghira@aii.pub.ro; [6] grigore.stamatescu@upb.ro
[2] Automatic Control and Systems Engineering Department, Politehnica University of Bucharest
 valentin.tanasa@acse.pub.ro

If the model of the process is known then nonlinear control strategies such as sliding mode control [1], backstepping control [7], passivity based control [8], [6] can be employed.

In the present paper it was chosen the backstepping strategy for nonlinear systems as the control method. The backstepping strategy is nowadays frequently used in the control design of nonlinear systems admitting strict feedback form. These specific state-space forms are used to model electro-mechanical systems and other sytems. In particular, a typical difficulty of these structures is that these exhibit a relative degree larger than one. This constitutes an obstacle in designing passivity based controllers and the backstepping procedure is a suitable tool to remove this obstacle[10].

The backstepping procedure, in its general formulation, gives the ingredients to compute controllers that stabilize the origin globally. Although this method can be compared with other state-feedback designs like dynamical linearization, its particularity is that the backstepping procedure takes into consideration the nonlinearity which can be useful for stabilization. It does not a priori cancel the nonlinearity since which can be included then into the controller. The main drawback of this procedure, highlighted when the number of cascade connections is large is that the controller expressions become quite complex. To cope with this, some relaxing procedures are proposed, which achieve semiglobal results such as high-gain designs or semiglobal backstepping [9].

## 2. Underlying Theory

In this article we consider a 2 sub-systems connection as follows

$$\dot{z}(t) = f(z(t)) + g(z(t))\zeta(t) \tag{1}$$

$$\dot{\zeta}(t) = f_a(z(t), \zeta(t)) + g_a(z(t), \zeta(t))u_c(t) \tag{2}$$

where the state $z$ and $\zeta$ are in $R^n$ and $R$ respectively and the control vector $u_c \in R$, $f$, $g$ and $f_a(\cdot), g(\cdot)$ in $R^n$ and in $R$ respectively.

The following result describes the stabilizing controller through the backstepping approach [10]

**Proposition 2.1** [5] - *Continuous-time backstepping - Consider the system (1)-(2), and suppose the existence of $\varphi(z)$ with $\varphi(0) = 0$ and a Lyapunov function $W(z)$, radially unbounded, such that*

$$\frac{\partial W}{\partial z}(f(z) + g(z)\varphi(z)) < 0, \forall z \in I \quad R^n / \{0\} \tag{3}$$

Then, if $g_a^{-1}(z, \zeta)$ exists for all $(z, \zeta)$, the state feedback control law

$$u_c = g_a(z,\zeta)^{-1}(\dot{\phi} - \frac{\partial W}{\partial z} g(z) - f_a(z,\zeta) - K_y(\zeta - \varphi)) \quad (4)$$

with $\dot{\Phi} = \frac{\partial f}{\partial z}(f(z) + g(z)\zeta)$ globally asymptotically stabilizes the origin of (1)-(2), with

$$V(z,\zeta) = W(z) + \frac{1}{2}(\zeta - \varphi(z))^2 \quad (5)$$

as a Lyapunov function.

   *Comments.* By considering $\varphi(z)$ as a fictitious control for the first z-dynamics, it follows from (3) that the fictitious state feedback $\zeta = \varphi(z)$ asymptotically stabilizes the dynamics (1) at the origin. Setting $y = \zeta - \varphi(z)$, (1)-(2) can be rewritten as

$$\dot{z}(t) = f(z) + g(z)(\varphi(z) + y) \quad (6)$$

$$\dot{y}(t) = f_a(z,\zeta) - \frac{\partial \varphi}{\partial z}\dot{z} + g_a(z,\zeta)u_c, \quad (7)$$

thus describing the second part as an y-error dynamics. By setting $V$ as in (5) and

$$u_c = g_a^{-1}(z,\zeta)(-f_a(z,\zeta) - y_0 + \frac{\partial \varphi}{\partial z}(f(z) + g(z)\zeta) + v) \quad (8)$$

one achieves $v \to y$ stabilisation with the Lyapunov function $V$ . Then, by setting $v = -K_y y$ with $K_y > 0$, one gets, (4) so that

$$\dot{V} = \frac{\partial W}{\partial z}(f(z) + g(z)\varphi) + \frac{\partial W}{\partial z} g(z)y + y^T(v - y_0)$$
$$= \frac{\partial W}{\partial z}(f(z) + g(z)\varphi) - K_y y^T y < 0 \quad (9)$$

because of (3). The global asymptotic stabilization at the origin follows, since $W(z)$ is radially unbounded.

 It follows that the backstepping procedure is very usefull for systems having a relative degree larger than one (e.g. the first dynamics $z$ is not directly driven by the control input). The speed of the convergence rate of the states to the equilibrium can be adapted in function of the choice of the gain values $K_y$ and $K_\varphi$. Larger values for $K_y$ and $K_\varphi$ means an increased speed and as consequence a reduced settling time. The drawback is that the amplitude of the control input is increased and this will be often saturated by the physical limitations. The right values of these parameters are chosen often experimentaly (or by means of simulations) and taking into account the trade off between speed and control saturation.

## 3. The experimental plant model and controller design

### 3.1 The plant description

The laboratory platform (ELWE Technick) used to test the proposed algorithms, available at Politehnica University Bucharest, consists of 3 main water tanks and a water reservoir. The flows of the water is assured by the means of 6 electro valves and 2 pomps and each liquid level is measured by one of the 3 piezoresistive transducers (VEGABAR 14).

In figure 1 is presented the schematic of a coupled-tank system that was configured on the laboratory platform. The water is pumped into tank $T_1$ and from there through a connection pipe (with a section area $S_1$) into the tank $T_3$. The water flows into reservoir through an electro valve (full open) with a section area $S_2$.

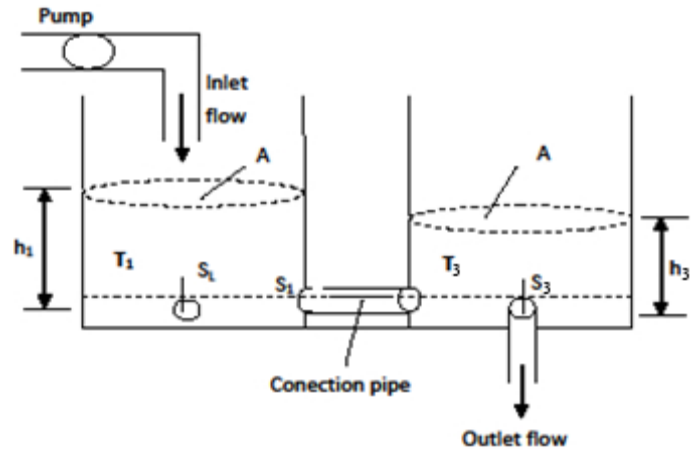

Figure 1: Schematic of coupled-tanks

The values of the plant parameters are described next:

- $A = 0.0154\ m^2$ - section area of each tank;
- $S_1 = S_3 = 5 \cdot 10^{-5}\ m^2$ - the section area of each pipeline;
- $Q_0 = 10^{-4}\ m^3 / s$ , $h_0 = 0.6\ m$ - the maximal values of the inlet flow and tank level respectively.

The connection of the plant with a PC is assured by 2 acquisition cards : "Humusoft MF624" for Pump 1 and level transducers and "National Instruments PCI-6503" which controls the electro valves. The real-time interface is configured in Simulink.

### 3.2 Dynamic model

The dynamic model of the coupled tanks is described by

$$\dot{h}_1 = \frac{1}{A} Q_i(t) - \frac{1}{A} c_1 \sqrt{|h_1 - h_3|} \, sign(h_1 - h_3)$$

$$\dot{h}_3 = \frac{1}{A} c_1 \sqrt{|h_1 - h_3|} \, sign(h_1 - h_3) - \frac{1}{A} c_2 \sqrt{h_3} \qquad (10)$$

with $c_1 = 1.0167 \cdot 10^{-4}$, $c_2 = 1.7253 \cdot 10^{-4}$ experimentally determined. The parameters $c_i = a z_i S_n \sqrt{2g}$ define the product between the flow coefficient $a z_i$, the pipe cross area and the gravitational acceleration . The input is the inlet volumetric flow rate $u(t) = Q_i(t)$, the state variables represents the liquid levels in tank $T_1$ and $T_3$ and the output is chosen equal to $h_1$. In the case of coupled tanks, the inequality $h_1 \geq h_3$ holds, in every operating point.

The structure of the system is similar with the one represented by equations (1)-(2) and a backstepping procedure can be applied. In a first step, the model has to be brought to an error representation by using the following change of coordinates:

$$z_1 = h_3, \ z_3 = \sqrt{h_1 - h_3} \qquad (11)$$

Also we have renamed the flow coefficients as follows: $a = \frac{c_1}{A}$; $b = \frac{c_2}{A}$. The input is divided into two components $u_s$ the stationary component and the actual control $v(t)$ respectively, hence $Q(t) = Q_0(u_s + v(t))$. The dynamics (10) are now rewritten as follows:

$$\dot{z}_1 = -b\sqrt{z_1} - a z_3; \dot{z}_3 = \frac{b}{2} \frac{\sqrt{z_1}}{z_3} - a + \frac{1}{2A z_3} Q_1 \qquad (12)$$

It can be noticed the strict-feedback structure of the proces is now available. Further we continue with a new change of coordinates in order to obtain the error model dynamics:

$$\eta_1 = \frac{z_1 - z_{1r}}{z_0} \ \ \eta_2 = \frac{z_3 - z_{3r}}{z_0} \qquad (13)$$

with $z_{3r} = \sqrt{h_{1red} - h_{3ref}}$. It follows next the error dynamic model.

$$\dot{\eta}_1 = -\frac{b}{z_0} \sqrt{z_0 \eta_1 + z_{1r}} - \frac{a}{z_0}(\eta_2 z_0 + z_{3r}) \qquad (14)$$

$$\dot{\eta}_2 = \frac{b}{2z_0} \frac{\sqrt{\eta_1 z_0 + z_{1r}}}{\eta_2 z_0 + z_{3r}} - \frac{a}{z_0} + \frac{Q_0}{2A z_0(\eta_2 z_0 + z_{3r})}(u_s + v(t)) \qquad (15)$$

### *3.3 Controller design*

In order to compute the controller by using the backstepping strategy in the same manner as it it given in Section 2, by looking to equations (1)-(2) and based on the error dynamics model we can identify:

$$f(\eta_1) = -\frac{b}{z_0}\sqrt{\eta_1 z_0 + z_{1r}} + \frac{a}{z_0}z_{3r}; \quad g = a; \quad g_a(\eta_1, \eta_2) = \frac{Q_0}{2Az_0}\frac{1}{z_0\eta_2 + z_{3r}}$$

$$f_a(\eta_1, \eta_2) = -\frac{b}{2z_0}\frac{\sqrt{\eta_1 z_0 + z_{1r}}}{z_0\eta_2 + z_{3r}} - \frac{a}{z_0} + \frac{Q_0}{2Az_0}\frac{u_s}{z_0\eta_2 + z_{3r}}; \quad (16)$$

### *3.3.1 The static control design*

As it was mentioned earlier $u_s$ is the static controller and its objectiv it to assure a zero steady state error. In order to find this component we impose that $\dot{\eta}_1 = 0$, $\dot{\eta}_2 = 0$ and $v = 0$ respectively. After some computations we get the following expressions:

$$u_s = \frac{aA}{Q_0}(z_0\eta_2 + z_{3r}) = \frac{aA}{Q_0}z_{3r} \quad (17)$$

### *3.3.2 The backstepping controller design*

We follow next the procedure proposed by Proposition 2.1. Let us define the initial Lyapunov function (of the first dynamics)

$$W(\eta_1) = \frac{1}{2}\eta_1^2 \quad (18)$$

Then there exists a function $\varphi(\eta_1)$ such that

$$\frac{\partial W}{\partial \eta_1}\big(f(\eta_1) + g\varphi(\eta_1)\big) < 0 \quad (19)$$

and this function can be computed as

$$\varphi(\eta_1) = \frac{1}{g_a(\eta_2)}\Big[\dot{\varphi}(\eta_1) - a\eta_1 - f_a(\eta_1, \eta_2) - K_y(\xi - \varphi(\eta_1))\Big] \quad (20)$$

with $\dot{\varphi}(\eta_1) = \frac{\partial \varphi}{\partial \eta_1}\dot{\eta}_1$. Then according to (8) the backstepping controller adapted to the problem at hand has the following structure

$$v = \frac{2Az_0(z_0\eta_2 + z_{3r})}{Q_0}\left(v_i - \eta_1 a - K_y y - v_{ii} + \frac{z}{z_0}\right) \quad (21)$$

with

$$v_i = \left(\frac{b}{2g}\sqrt{1}\sqrt{z_0\eta_1 + z_{1r}} - \frac{K_\varphi}{g}\right)(ay - K_\varphi\eta_1)$$

$$v_{ii} = \frac{1}{2z_0\sqrt{z_0\eta_1 + z_{1r}}}(b\sqrt{z_0\eta_1 + z_{1r}} + az_{3r}) \text{ and } y = \eta_2 - \varphi(\eta_1)$$

This controller assures the stabilization of the process (defined by the error model) in the point where $(\eta_1, \eta_2) = (0,0)$. In the way these variables were defined it follows that the $0$ equilibrium correspond to the case when the levels in the tanks have riched their setpoints values. The stability of this controller can be proved by verifying that the extended Lyapunov function $V(\eta_1, \eta_2) = W(\eta_1) + \frac{1}{2}y^2$ has its derivative negative for all admissible values of the state variables $\eta_1$ and $\eta_2$. The controller proposed guarantees the stabilization of the system around the points given by the references.

## 4. Results

The designed controller has been tested on the experimental platform described in Section 3. The controller has been implemented in Simulink and tested by using a real-time configuration. The sampling period has been set to 1 seconds (small enough to ensure a good emulation of the continuous time controller).

Simulations have been carried out by setting the reference of the level in the first tank to 80% and 50% respectively (blue line in Figure 2). The evolutions of the level in the two tanks are illustrated in the next figures with lines red and light blue. The evolution of the control input is represented with green-line. The control structure was tested in normal conditions with no major perturbations respectively when some faults were introduced (see Figure 3).

It must be noticed that in steady state, the level in the first tank follows very accurate the reference value. As the tanks are coupled, the level in the second tank has the same characteristic. In Figure 3 it can be observed that the controller reacts rapidly at the changes in the process characteristiques in the direction to counteract the faults that occured.

## 5. Conclusions

In this article a backstepping controller has been designed to deal with the control of liquid level in a coupled tank system. The proposed strategy is a non-linear control law which avoids the clasical liniarization designs when dealing with linear systems. Also this control law offers the same performances in any operating point. The results confirm the efficacy of this controller and some steady state errors can be observed due to the unmodeled dynamics or non-linearities. The procedure proposed requires that the model of the process to be exactly determined.

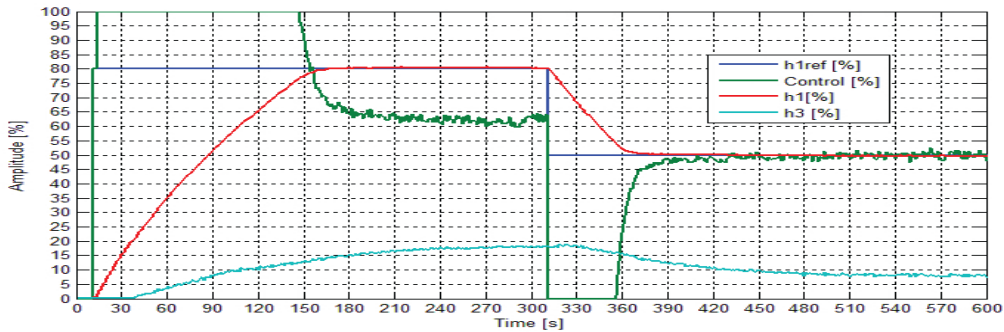

Figure 2: Simulation results without perturbation

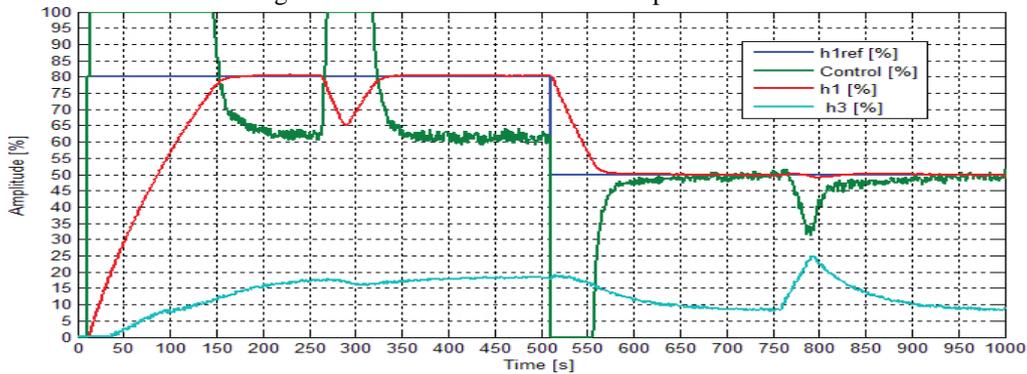

Figure 3: Simulation results with perturbation

## R E F E R E N C E S